\documentclass[aps,prl,twocolumn,amsfonts,amssymb,amsmath,floatfix,showpacs]{revtex4}
\usepackage[dvips]{graphicx}

\renewcommand{\@}{\partial}
\newcommand{\const}{\mathrm{const}}
\renewcommand{\d}{\mathrm{d}}
\newcommand{\delay}{\Delta}
\newcommand{\Dirac}{\delta}
\newcommand{\Heav}{\Theta}
\newcommand{\Irh}{I_{\mathrm{rh}}}
\newcommand{\Ist}{I_s}
\renewcommand{\L}{\mathcal{L}}
\newcommand{\lambdac}{\lambda_c}
\newcommand{\Real}{\mathbb{R}}
\newcommand{\sech}{\mathop{\mathrm{sech}}\nolimits}
\newcommand{\stept}{h_t}
\newcommand{\stepx}{h_x}
\newcommand{\tst}{t_s}
\newcommand{\thresh}{\theta}
\newcommand{\ust}{u_s}
\newcommand{\ucr}{u_*}
\newcommand{\ustcr}{\ust^*}
\newcommand{\xst}{x_s}

\newcommand{\eg}{e.g.}
\newcommand{\ie}{i.e.}

\newcommand{\eqlabel}[1]{\label{eq:#1}}
\def\eq(#1){(\ref{eq:#1})}
\def\eqtwo(#1,#2){(\ref{eq:#1},\ref{eq:#2})}
\def\eqthree(#1,#2,#3){(\ref{eq:#1},\ref{eq:#2},\ref{eq:#3})}

\newcommand{\dblfigure}[3]{\begin{figure*}[ht!]#1\caption[]{#2}\label{fig:#3}\end{figure*}}
\newcommand{\sglfigure}[3]{\begin{figure}[ht!]#1\caption[]{#2}\label{fig:#3}\end{figure}}
\newcommand{\Fig}[1]{Fig.~\ref{fig:#1}}
\newcommand{\fig}[1]{Fig.~\ref{fig:#1}}

\begin{document}
\title{An analytical approach to initiation of propagating fronts}
\author{I. Idris}
\author{V. N. Biktashev}
\affiliation{Department of Mathematical Sciences, 
  University of Liverpool, Liverpool L69 7ZL, UK }
\date{\today}
\begin{abstract}
  We consider the problem of initiation of a propagating wave
  in a one-dimensional excitable fibre. 
  In the Zeldovich-Frank-Kamenetsky equation, a.k.a. Nagumo equation, the key role is played by the
  ``critical nucleus'' solution whose stable
  manifold is the threshold surface separating initial conditions
  leading to initiation of propagation and to decay.
  Approximation of this manifold by its tangent linear space yields an analytical criterion
  of initiation which is in a good agreement with direct numerical simulations. 
\end{abstract}
\maketitle

Threshold phenomena are widespread in bistable dissipative systems.
If such a system is spatially extended, then fronts switching from one
local state to the other can propagate. Propagating fronts, or trigger
waves, play important roles in such diverse physical situations as
self-heating in metals and superconductors, phase transitions,
combustion and other chemical reaction waves, and biological
signalling systems~\cite{ Gurevich-Mints-1987,%
  Cross-Hohenberg-1993,%
  Keener-Sneyd-1998,%
  Merzhanov-Rumanov-1999,%
  Bychkov-Liberman-2000,%
  Murray-2002%
} to name a few. In biology and chemistry they often appear as a fast
stage of pulse waves in ``excitable systems'' \cite{%
  Keener-Sneyd-1998,%
  Murray-2002,%
  Krinsky-Swinney-1991,%
  Holden-etal-1991%
}. The question of existence of such waves in particular mathematical
models is well studied. However, whether a propagating wave will
actually be observed depends on initial conditions.  Understanding
conditions of initiation of propagating fronts or pulses is very
important in applications. In heart such waves trigger
coordinated contraction of the muscle and failure of initiation can
cause or contribute to serious or fatal medical conditions, or render
inefficient the work of pacemakers or defibrillators
\cite{Zipes-Jalife-2004}. 
In combustion, 
understanding of initiation is of critical importance
for safety in storage and transport of combustible materials
\cite{Shah-etal-2008}.

Mathematically, after the external initiating stimulus has finished, the problem is
reduced to classification of initial conditions that will or will not
lead to a propagating wave solution. This problem is difficult as it
is essentially non-stationary, spatially extended and nonlinear, and
does not have any helpful symmetries. Yet the problem is so important
that analytical answers are highly desirable even if not very
accurate.

Early attempts of analytical treatment of the initiation problems, including
the spatially extended ones, used linear description supplemented with heuristic conditions
to represent the threshold \cite{%
  Lapicque-1907,%
  Blair-1932,%
  Hill-1936,%
  Rushton-1937,%
  Noble-1972%
} and, more recently, low-dimensional Galerkin style approximations of
the partial differential equations \cite{%
  Suarez-Sicardi-1996,%
  Neu-etal-1997%
}

In the last two decades, this problem has been analysed 
from the dynamical systems theory viewpoint~\cite{
  McKean-Moll-1985,
  Flores-1989,
  Flores-1991,
  Moll-Rosencrans-1990,
  Suarez-Sicardi-1996,%
  Neu-etal-1997,
  Idris-Biktashev-2007
}. These studies identified the importance of certain ``critical
solutions'', whose codimension-1 (center-)stable manifold acts as the critical
surface separating the basins of attraction of initiation and decay.
This understanding was used in sophisticated numerical methods of calculating initiation thresholds,
\eg\ \cite{Moll-Rosencrans-1990}.

Here we propose a practical method of defining the initiation criteria
analytically. The idea is based on the linearization of the
(center-)stable manifold of the critical solution by its linear
tangent, the (center-)stable space. One would expect that this should
work well for initial conditions sufficiently close to the critical
nucleus.  However, how close it should be to give a reasonable
approximation is not clear \textit{a priori}. We consider a test case
with very crude initial conditions, in the form of rectangular pulses,
and the analytical criterion gives surprisingly good agreement with
direct numerical simulations.

\dblfigure{
  \includegraphics{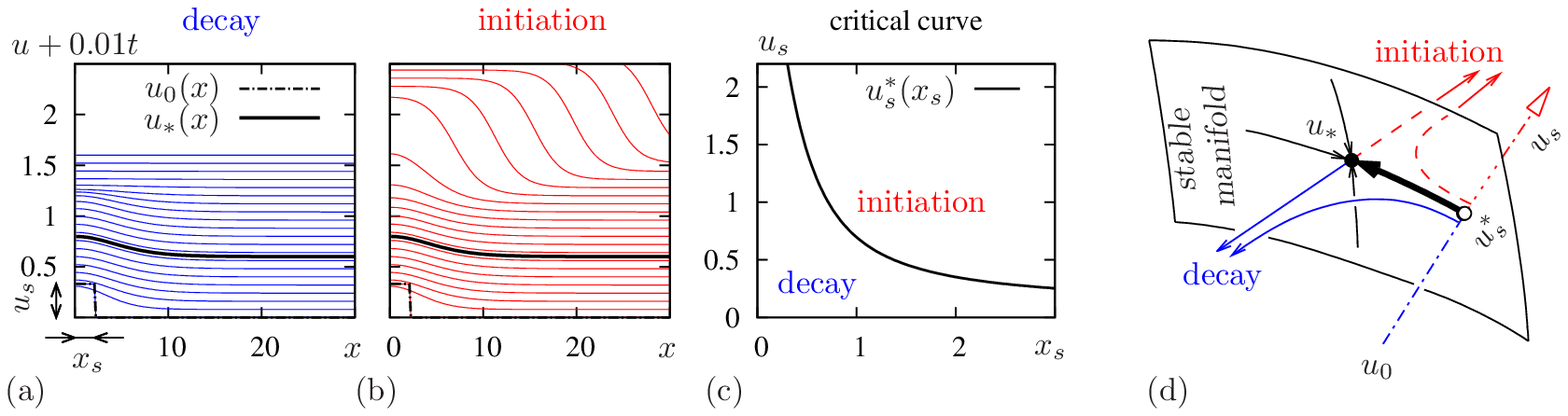}
}{
  (color online)
  (a,b) Response to an below- and above-threshold initial perturbation in ZFK equation,
  \eqthree(RDE,fcubic,rect). 
  Parameter values: $\theta=0.13$, $\Ist=0$, $\xst=2.10$ for both, subthreshold 
  $\ust=0.3304831$ (a) and superthreshold 
  $\ust=0.3304833$ (b) cases, 
  numerics using central difference centered in space with step $\stepx=0.15$
  and forward Euler in time with step $\stept=0.01$. 
  Dash-dotted black lines: initial conditions, 
  bold solid black lines: the critical nuclei.
  (c) The corresponding critical strength-extent curve,
  separating initiation initial conditions from decay initial conditions.
  (d)
  The sketch of a stable manifold of the critical solution for the ZFK
  equation.  The critical nucleus is represented by the black dot; the
  critical trajectories, constituting the stable manifold, are shown in
  black.  The family of initial conditions is represented by the
  dash-dotted line.  The bold black line is the critical trajectory with
  initial condition in that family.  The sub-threshold trajectories are
  represented by the blue line, while the red lines represent
  super-threshold trajectories. Note that the point where the initial
  condition intersect the stable manifold is shown as the empty circle.
}{fig1}

We consider a one-component reaction-diffusion equation 
\begin{equation}
 u_t = u_{xx} + f(u), \eqlabel{RDE}
\end{equation}
with bistable kinetics $f(u)$. As an archetypical example, we consider
Zeldovich-Frank-Kamenetsky equation suggested to describe flame
propagation~\cite{ZFK-1938}, which is also known
as Nagumo equation in its capacity as
the fast equation in
the FitzHugh-Nagumo system, suggested as a simplified model of nerve
conduction~\cite{FitzHugh-1961,Nagumo-etal-1962}.  This
equation has the kinetics in the form
\begin{equation}
f(u)=u(u-\thresh)(1-u), \qquad \thresh=\const<1/2.  \eqlabel{fcubic}
\end{equation}
Equations \eq(RDE) have propagating front solutions, 
\[
  u=U(x-ct-\delay), \quad \delay=\const      
\]
\eg\ for \eq(fcubic), 
\[
  U(\xi) = \frac{1}{1+e^{\xi/\sqrt2}}, 
  \quad  
  c=\frac{1-2\thresh}{\sqrt{2}}.              
\]
We consider a half-infinite cable which is driven away at $t=0$ from the resting state $u=0$
by an instantaneous stimulus of amplitude $\ust$ and spatial extent $\xst$ at $t=0$ and/or by
a current injection at $x=0$ of amplitude $\Ist$ lasting for time $\tst$,
\begin{eqnarray}
&& u_t = u_{xx} + f(u), \quad (x,t)\in \Real^+ \times \Real^+, \eqlabel{eq+domain} 
\\
&& u_x(0,t) =  \Ist g(t;\tst), \quad g(+\infty;\tst)=0 , \eqlabel{bc} 
\\
&& u(x,0) = \ust h(x;\xst), \quad h(+\infty;\xst)=0 , \eqlabel{ic}
\end{eqnarray}
or equivalently
\begin{eqnarray}
&& u_t = u_{xx} + f(u) + 2\Ist g(t;\tst) \, \Dirac(x), \; 
  \nonumber\\&& \qquad\qquad 
(x,t)\in \Real \times \Real^+; \eqlabel{eq+domain-even} 
\\
&& u(x,0) = \left\{
  \begin{array}[c]{l}
    \ust h(x;\xst), \quad x\ge0,  \\
    \ust h(-x;\xst), \quad x<0.  
    
  \end{array}\right.                        \eqlabel{ic-even}
\end{eqnarray}
where $\Dirac()$ is the Dirac delta function
(generalization for a generic stimulus $\Ist g(x,t)$ is straighforward). 
Specifically, we consider stimuli of rectangular profiles, 
\begin{equation}
  g(t;\tst)=\Heav(\tst-t), \qquad h(x;\xst)=\Heav(\xst-x),  \eqlabel{rect}
\end{equation}
where $\Heav()$ is the Heaviside step function. 

Depending on parameters $\ust$, $\xst$, $\Ist$ and $\tst$, problem
\eqthree(eq+domain,bc,ic) can typically produce either a ``decay''
solution such that $\max_{x}u(x,t)\to0$, $t\to\infty$, see
\fig{fig1}(a), or an ``initiation'' solution such that
$\max_{x}|u(x,t)-U(x-ct-\delay)|\to0$, $t\to\infty$ for some
$\delay\in\Real$, see \fig{fig1}(b).  Naturally, in the even extension
\eqtwo(eq+domain-even,ic-even), the ``initiation'' solution
produces two fronts propagating both ways. Our goal is a
condition that would predict which of the two outcomes will take place
for given $\ust$, $\xst$, $\Ist$ and $\tst$.  The curve in the
$(\tst,\Ist)$ plane, at $\ust=0$, separating the two outcomes, is
widely known as the strength-duration curve. We will also consider a
similar critical curve in the $(\xst,\ust)$ plane at $\Ist=0$, see
\fig{fig1}(c), which we will call strength-extent curve.

We consider first the case $\Ist=0$, and following~\cite{Flores-1989}, 
review the fundamental role of the critical
nucleus solution $\ucr(x)$, which is defined as a nontrivial stationary
solution of \eq(RDE), \ie\
\[
  \ucr'' + f(\ucr)=0, \qquad \ucr(x)\ne\const, 
\]
\eg\ for \eq(fcubic), 
\[
  \ucr(x) = \frac{3\thresh\sqrt{2}}{(1+\thresh)\sqrt{2}+\cosh(x\sqrt{2}) \sqrt{2-5\thresh+\thresh^2}}. 
\]
It is then demonstrated that such a solution is unstable. Consider linearization of \eq(RDE) near it, 
\(
  u(x,t)=\ucr(x)+v(x,t), \qquad v(x,t)\ll1,            
\)
then
\(
  v_t=\L v,                                           
\)
where
\(
   \L=\@_x^2 + f_u(\ucr(x)) . 
\)
Stability of $\ucr$ is determined by the spectrum of $\L$,
\begin{equation}
  \L \phi_j = \lambda_j \phi_j .           \eqlabel{ev}
\end{equation}
Since $\L$ is a Sturm-Liouville operator, all its eigenvalues $\lambda_j$ are real. 

Notice that $\L \@_x \ucr = 0$
so $\ucr'(x)$ is an eigenfunction corresponding to eigenvalue 0.
By Sturm's oscillation theorem, if eigenvalues of the discrete spectrum
are ordered so that
\( 
  \lambda_1>\lambda_2>\lambda_3>\dots 
\)
then egenfunction $\phi_j$ shall have precisely $j-1$ zeros. 
The critical nucleus $\ucr(x)$ is an even function and has a single maximum at $x=0$,
so $\ucr'(x)$ has
exactly one zero, and therefore we have
\[
  C_2 \ucr'(x)=\phi_2, \qquad \lambda_2=0,
\]
for some $C_2\ne0$.
This implies that $\ucr$ is unstable, and there is exactly one positive
eigenvalue, $\lambda_1>0$, with a corresponding $\phi_1(x)>0$.  The
continuous spectrum of $\L$ is $\{\lambda\}=(-\infty,\lambdac]$, where
$\lambdac=\lim\limits_{x\to\pm\infty}\left[\@_uf(u)\right]_{u=\ucr(x)}=f'(0)<0$.
Hence in the phase space of \eq(eq+domain-even), equilibrium $\ucr$ is a saddle
point, with only one unstable direction. Its stable manifold~\endnote{
  It is a stable manifold in \eq(eq+domain) and in the even-function
  subspace of \eq(eq+domain-even), and a center-stable manifold in the
  full phase space of \eq(eq+domain-even).
} has
therefore codimension one and, as such, partitions the phase space. One
part of the phase space corresponds to the decay solutions, and the
other to the initiation solutions (\fig{fig1}(d)).  A one-parametric
family of initial conditions \eq(ic), say with a fixed
$\xst>0$ and the parameter $\ust$, will cross the stable manifold once, 
say at $\ust=\ustcr(\xst)$.  For $\ust<\ustcr(\xst)$, we have
decay, and for $\ust>\ustcr(\xst)$ 
initiation. This defines the strength-extent curve $\ust=\ustcr(\xst)$.
The role of the stable manifold of the critical nucleus $\ucr$ as the
threshold surface in the phase space is an empirically verifiable
fact: it means that the critical nucleus will be observed as a
transient for any initial conditions sufficiently close to the
threshold (see \fig{fig1} (a,b) for $t\lesssim100$).

Now we shall use this understanding to construct an analytical
criterion of initiation. Our idea is to replace the stable manifold of
$\ucr$ by its tangent, \ie\ the stable space. This implies considering
the initiation problem in the linear approximation around $\ucr(x)$.
Continuing with the case $\Ist=0$, we get
\[
  u(x,t) = \ucr(x) + \sum\limits_{j=1}^{\infty} a_j e^{\lambda_jt} \phi_j(x), 
\]
where for brevity the summation is assumed both over the discrete and the continuous spectrum.
If we choose the eigenfunctions $\phi_j(x)$ normalized, then
\( 
  a_j = \int\limits_{-\infty}^{\infty} \phi_j(x) \bigg( u(x,0) - \ucr(x) \bigg) \,\d{x} .
\)
Eigenfunction $\phi_2(x)=\ucr'(x)$ is odd, $\ucr(x)$ and $u(x,0)$ are even, hence
$a_2=0$, and
$\sum\limits_{j=3}^{\infty}a_je^{\lambda_jt}\phi_j(x)\to0$ as
$t\to\infty$ since $\lambda_j\le\lambda_3<0$ for $j\ge3$. Hence in
this approximation $u(x,t)\to \ucr(x)$ if and only if $a_1=0$. So the equation of
the stable space, which is an approximation of the critical manifold,
is $a_1=0$ or
\begin{equation}
  \int\limits_{0}^{\infty} \phi_1(x) \left(\ust h(x;\xst) - \ucr(x) \right) \,\d{x} = 0 .   \eqlabel{critcurve}
\end{equation}
This is a finite equation for $\xst$, $\ust$, which provides the
desired analytical definition of the strength-extent curve.

For $\Ist\ne0$, we have 
\[
  u(x,t) = \ucr(x) + \sum\limits_{j=1}^{\infty} A_j(t) \, \phi_j(x),
\]
where
\(
  A_j(0) = \int\limits_{-\infty}^{\infty} \phi_j(x) \bigg( u(x,0) - \ucr(x) \bigg) \,\d{x}
\)
and
\(
  \d{A_j}/\d{t} = \lambda_j A_j + 2 \Ist \, g(t) \, \phi_j(0),            
\)
which can be solved in quadratures for a given $g(t)$, and then the critical condition is
$ A_1(+\infty) = 0$, or 
\begin{equation}
  A_1(0) + 2 \Ist \phi_1(0) \int\limits_{0}^{\infty} e^{-\lambda_1t} g(t) \, \d{t} = 0 .
                                                               \eqlabel{strength-duration}
\end{equation}

Now we consider an example with explicit answers.
For \eq(fcubic), if $\thresh\ll1$, then $\ucr=\O{\thresh}$, and 
as in \cite{Neu-etal-1997}, for
$u\lesssim\thresh$ we can approximate
\begin{equation}
  f(u) \approx u(u-\thresh)                                   \eqlabel{fquad}
\end{equation}
and then
\(
  \ucr \approx \frac{3}{2} \thresh \sech^2\left(x\sqrt{\thresh}/2\right). 
\)
In this approximation, to solve the eigenvalue problem \eq(ev), 
it is convenient to change variables $v(x)=\psi(z)$, 
$z=\tanh(x\sqrt{\theta}/2)$, then 
\[
\left( (1-z^2)\psi'\right)' + \left( 12 - \frac{4(1 + \lambda/\theta)}{1 - z^2} \right)\psi = 0,
\quad
\psi(\pm 1) = 0, \label{eq:SSLP-zfk}
\]
solutions of which are associated Legendre functions \cite{Gradshteyn-Ryzhik2000}.
In particular, we find that 
\[
  \lambda_1=5\thresh/4, \quad \phi_1(x) = C_1 \sech^3(x\sqrt{\thresh}/2)
\]
for some $C_1\ne0$.

For $\Ist=0$ and $h(x;\xst)=\Heav(\xst-x)$, equation \eq(critcurve) then gives
an explicit equation for the strength-extent curve
\begin{eqnarray}
\ust &=& 
  \frac{9\thresh}{8}\left[ \frac{2}{\pi}\tanh\left(\frac{\xst\sqrt{\thresh}}{2}\right)
  \sech\left(\frac{\xst\sqrt{\thresh}}{2}\right)
\right. \nonumber\\ && \left.\qquad\mbox{} 
+\frac{4}{\pi}\arctan\left(e^{\xst\sqrt{\thresh}/2}\right)-1 \right]^{-1} .
                                                                        \eqlabel{xu-quad}
\end{eqnarray}

For $\ust=0$ and $g(t;\tst)=\Heav(\tst-t)$,  we have
$A_1(0)=\frac98 \pi\sqrt{\thresh} C_1$
and equation \eq(strength-duration) gives the classical
Lapicque-Blair-Hill~\cite{Lapicque-1907,Blair-1932,Hill-1936}
equation for the strength-duration curve,
\begin{equation}
  \Ist=\frac{\Irh}{1-e^{-\tst/\tau}},                                    \eqlabel{LBH}
\end{equation} 
with rheobase
\begin{equation}
  \Irh=\frac{ \phi_1(0) }%
  {\lambda_1\int_{0}^{\infty}\phi_1(x)\ucr(x)\,\d{x}}
  =\frac{45}{64}\pi\theta^{3/2}                                         \eqlabel{rheobase}
\end{equation} 
and chronaxie
\begin{equation}
  \tau= \left( \lambda_1 \right)^{-1} =\frac{4}{5\theta}.               \eqlabel{chronaxie}
\end{equation}

\sglfigure{
  \includegraphics{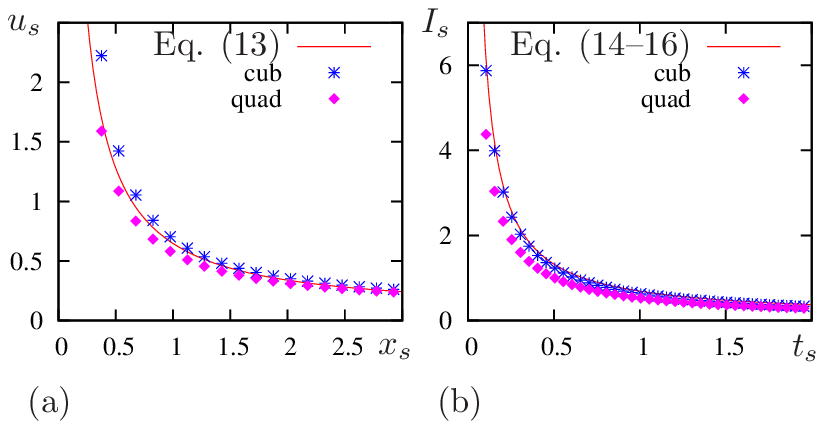}
}{
  (color online)
  Comparison of analytical predictions with numerical simulations. 
  (a) Strength-extent curves for rectangular initial conditions. 
  (b) Strength-duration curves for point stimulation. 
  Red solid lines: analytical approximations, \eq(xu-quad) for (a) and \eqthree(LBH,rheobase,chronaxie) for (b). 
  Blue stars (``cub''): numerical results for cubic kinetics \eq(fcubic). 
  Magenta diamonds (``quad''): numerical results for quadratic kinetics \eq(fquad).
}{fig2}

\Fig{fig2} illustrates the quality of the analytical critical curves
\eq(xu-quad) and \eqthree(LBH,rheobase,chronaxie), both compared to the
curves obtained by direct numerical simulations
for the quadratic nonlinearity \eq(fquad) valid for small
$\thresh$, and the original cubic nonlinearity \eq(fcubic). For the
chosen parameter values, the error introduced by linear approximation
of the stable manifold of the critical nucleus is of the same order of
magnitude as the error introduced by the quadratic approximation of
the nonlinearity.

In conclusion, we have obtained analytical expressions for initiation
criteria for a concrete simple example. Such criteria were obtained
experimentally and numerically and any analytical expression was
through fitting; we have deduced it mathematically \textit{ab initio},
via a clearly defined procedure. The expressions are simple enough to
be useful in practice, but the procedure of obtaining them is probably
more important as it can be extended to other models. The expression
for the strength-extent curve is specific for the ZFK equation and
will have a different form for a different model. However, the
temporal strength-duration curve is universal, up to the values of two
constants, and it coincides precisely with a classical form used for
over 100 years for analytical fitting of empirical data.

The general principle, linear approximation of the (center-)stable
manifold of the critical solution, easily admits extensions, \eg\ for
different temporal and spatial profiles of the initiation stimuli,
different initiation protocols, possibility of optimization, say with
respect to the total energy required to initiate a wave etc.

It also can be extended to other threshold systems, whenever the
critical solution can be identified, including those having critical
solutions which are not critical nuclei 
\cite{Idris-Biktashev-2007}. In such systems, an
additional problem is anticipated, as one cannot use the even
$(x\to-x)$ embedding and have to take into account the translational
symmetry of the problem posed on the whole real axis.

Authors are grateful to J.~Brindley for inspiring discussions and
R. Su{\'a}rez Antola for bibliographic advice.  The study has been
supported in part by EPSRC grant GR/S75314/01 and MacArthur Foundation
grant 71356-01.


\begin{thebibliography}{26}
\expandafter\ifx\csname natexlab\endcsname\relax\def\natexlab#1{#1}\fi
\expandafter\ifx\csname bibnamefont\endcsname\relax
  \def\bibnamefont#1{#1}\fi
\expandafter\ifx\csname bibfnamefont\endcsname\relax
  \def\bibfnamefont#1{#1}\fi
\expandafter\ifx\csname citenamefont\endcsname\relax
  \def\citenamefont#1{#1}\fi
\expandafter\ifx\csname url\endcsname\relax
  \def\url#1{\texttt{#1}}\fi
\expandafter\ifx\csname urlprefix\endcsname\relax\def\urlprefix{URL }\fi
\providecommand{\bibinfo}[2]{#2}
\providecommand{\eprint}[2][]{\url{#2}}

\bibitem[{\citenamefont{Gurevich and Mints}(1987)}]{Gurevich-Mints-1987}
\bibinfo{author}{\bibfnamefont{A.~V.} \bibnamefont{Gurevich}} \bibnamefont{and}
  \bibinfo{author}{\bibfnamefont{R.~G.} \bibnamefont{Mints}},
  \bibinfo{journal}{Rev. Mod. Phys.} \textbf{\bibinfo{volume}{59}},
  \bibinfo{pages}{941} (\bibinfo{year}{1987}).

\bibitem[{\citenamefont{Cross and Hohenberg}(1993)}]{Cross-Hohenberg-1993}
\bibinfo{author}{\bibfnamefont{M.~C.} \bibnamefont{Cross}} \bibnamefont{and}
  \bibinfo{author}{\bibfnamefont{P.~C.} \bibnamefont{Hohenberg}},
  \bibinfo{journal}{Rev. Mod. Phys.} \textbf{\bibinfo{volume}{65}},
  \bibinfo{pages}{851} (\bibinfo{year}{1993}).

\bibitem[{\citenamefont{Keener and Sneyd}(1998)}]{Keener-Sneyd-1998}
\bibinfo{author}{\bibfnamefont{J.~P.} \bibnamefont{Keener}} \bibnamefont{and}
  \bibinfo{author}{\bibfnamefont{J.}~\bibnamefont{Sneyd}},
  \emph{\bibinfo{title}{Mathematical Physiology}}
  (\bibinfo{publisher}{Springer}, \bibinfo{year}{1998}).

\bibitem[{\citenamefont{Merzhanov and Rumanov}(1999)}]{Merzhanov-Rumanov-1999}
\bibinfo{author}{\bibfnamefont{A.~G.} \bibnamefont{Merzhanov}}
  \bibnamefont{and} \bibinfo{author}{\bibfnamefont{E.~N.}
  \bibnamefont{Rumanov}}, \bibinfo{journal}{Rev. Mod. Phys.}
  \textbf{\bibinfo{volume}{71}}, \bibinfo{pages}{1173} (\bibinfo{year}{1999}).

\bibitem[{\citenamefont{Bychkov and Liberman}(2000)}]{Bychkov-Liberman-2000}
\bibinfo{author}{\bibfnamefont{V.~V.} \bibnamefont{Bychkov}} \bibnamefont{and}
  \bibinfo{author}{\bibfnamefont{M.~A.} \bibnamefont{Liberman}},
  \bibinfo{journal}{Physics Reports} \textbf{\bibinfo{volume}{325}},
  \bibinfo{pages}{115} (\bibinfo{year}{2000}).

\bibitem[{\citenamefont{Murray}(2002)}]{Murray-2002}
\bibinfo{author}{\bibfnamefont{J.~D.} \bibnamefont{Murray}},
  \emph{\bibinfo{title}{Mathematical Biology: {I.} An Introduction}}
  (\bibinfo{publisher}{Springer}, \bibinfo{year}{2002}).

\bibitem[{\citenamefont{Krinsky and Swinney}(1991)}]{Krinsky-Swinney-1991}
\bibinfo{editor}{\bibfnamefont{V.}~\bibnamefont{Krinsky}} \bibnamefont{and}
  \bibinfo{editor}{\bibfnamefont{H.}~\bibnamefont{Swinney}}, eds.,
  \emph{\bibinfo{title}{Wave and patterns in biological and chemical excitable
  media}} (\bibinfo{publisher}{North-Holland}, \bibinfo{address}{Amsterdam},
  \bibinfo{year}{1991}).

\bibitem[{\citenamefont{Holden et~al.}(1991)\citenamefont{Holden, Markus, and
  Othmer}}]{Holden-etal-1991}
\bibinfo{editor}{\bibfnamefont{A.~V.} \bibnamefont{Holden}},
  \bibinfo{editor}{\bibfnamefont{M.}~\bibnamefont{Markus}}, \bibnamefont{and}
  \bibinfo{editor}{\bibfnamefont{H.~G.} \bibnamefont{Othmer}}, eds.,
  \emph{\bibinfo{title}{Nonlinear wave processes in excitable media}}
  (\bibinfo{publisher}{Plenum Press}, \bibinfo{year}{1991}).

\bibitem[{\citenamefont{Zipes and Jalife}(2004)}]{Zipes-Jalife-2004}
\bibinfo{editor}{\bibfnamefont{D.~P.} \bibnamefont{Zipes}} \bibnamefont{and}
  \bibinfo{editor}{\bibfnamefont{J.}~\bibnamefont{Jalife}}, eds.,
  \emph{\bibinfo{title}{Cardiac electrophysiology: From cell to bedside}}
  (\bibinfo{publisher}{W B Saunders Co}, \bibinfo{year}{2004}).

\bibitem[{\citenamefont{Shah et~al.}(2008)\citenamefont{Shah, Brindley,
  McIntosh, and Rademacher}}]{Shah-etal-2008}
\bibinfo{author}{\bibfnamefont{A.~A.} \bibnamefont{Shah}},
  \bibinfo{author}{\bibfnamefont{J.}~\bibnamefont{Brindley}},
  \bibinfo{author}{\bibfnamefont{A.~C.} \bibnamefont{McIntosh}},
  \bibnamefont{and}
  \bibinfo{author}{\bibfnamefont{J.}~\bibnamefont{Rademacher}},
  \bibinfo{journal}{Nonlinear Analysis: Real World Applications}
  \textbf{\bibinfo{volume}{9}}, \bibinfo{pages}{562} (\bibinfo{year}{2008}).

\bibitem[{\citenamefont{Lapicque}(1907)}]{Lapicque-1907}
\bibinfo{author}{\bibfnamefont{L.}~\bibnamefont{Lapicque}}, \bibinfo{journal}{J
  Physiol (Paris)}  (\bibinfo{year}{1907}).

\bibitem[{\citenamefont{Blair}(1932)}]{Blair-1932}
\bibinfo{author}{\bibfnamefont{H.~A.} \bibnamefont{Blair}}, \bibinfo{journal}{J
  Gen Physiol} \textbf{\bibinfo{volume}{15}}, \bibinfo{pages}{709}
  (\bibinfo{year}{1932}).

\bibitem[{\citenamefont{Hill}(1936)}]{Hill-1936}
\bibinfo{author}{\bibfnamefont{A.~V.} \bibnamefont{Hill}},
  \bibinfo{journal}{Proc R Soc Lond (Biol)} \textbf{\bibinfo{volume}{119}},
  \bibinfo{pages}{305} (\bibinfo{year}{1936}).

\bibitem[{\citenamefont{Rushton}(1937)}]{Rushton-1937}
\bibinfo{author}{\bibfnamefont{W.}~\bibnamefont{Rushton}},
  \bibinfo{journal}{Proc. Roy. Soc. Lond. ser. B}
  \textbf{\bibinfo{volume}{124}}, \bibinfo{pages}{210} (\bibinfo{year}{1937}).

\bibitem[{\citenamefont{Noble}(1972)}]{Noble-1972}
\bibinfo{author}{\bibfnamefont{D.}~\bibnamefont{Noble}}, \bibinfo{journal}{J.
  Physiol.} \textbf{\bibinfo{volume}{226}}, \bibinfo{pages}{573}
  (\bibinfo{year}{1972}).

\bibitem[{\citenamefont{Suarez~Antola and
  Sicardi~Schifino}(1996)}]{Suarez-Sicardi-1996}
\bibinfo{author}{\bibfnamefont{R.~E.} \bibnamefont{Suarez~Antola}}
  \bibnamefont{and} \bibinfo{author}{\bibfnamefont{A.~C.}
  \bibnamefont{Sicardi~Schifino}}, \bibinfo{journal}{Physica D}
  \textbf{\bibinfo{volume}{89}}, \bibinfo{pages}{427} (\bibinfo{year}{1996}).

\bibitem[{\citenamefont{Neu et~al.}(1997)\citenamefont{Neu, Preissig, and
  Krassowska}}]{Neu-etal-1997}
\bibinfo{author}{\bibfnamefont{J.~C.} \bibnamefont{Neu}},
  \bibinfo{author}{\bibfnamefont{R.~S.} \bibnamefont{Preissig}},
  \bibnamefont{and}
  \bibinfo{author}{\bibfnamefont{W.}~\bibnamefont{Krassowska}},
  \bibinfo{journal}{Physica D} \textbf{\bibinfo{volume}{102}},
  \bibinfo{pages}{285} (\bibinfo{year}{1997}).

\bibitem[{\citenamefont{McKean and Moll}(1985)}]{McKean-Moll-1985}
\bibinfo{author}{\bibfnamefont{H.~P.} \bibnamefont{McKean}} \bibnamefont{and}
  \bibinfo{author}{\bibfnamefont{V.}~\bibnamefont{Moll}},
  \bibinfo{journal}{Bull. AMS} \textbf{\bibinfo{volume}{12}},
  \bibinfo{pages}{255} (\bibinfo{year}{1985}).

\bibitem[{\citenamefont{Flores}(1989)}]{Flores-1989}
\bibinfo{author}{\bibfnamefont{G.}~\bibnamefont{Flores}}, \bibinfo{journal}{J.
  Diff. Eq.} \textbf{\bibinfo{volume}{80}}, \bibinfo{pages}{306}
  (\bibinfo{year}{1989}).

\bibitem[{\citenamefont{Flores}(1991)}]{Flores-1991}
\bibinfo{author}{\bibfnamefont{G.}~\bibnamefont{Flores}},
  \bibinfo{journal}{SIAM J. Math. Anal.} \textbf{\bibinfo{volume}{22}},
  \bibinfo{pages}{392} (\bibinfo{year}{1991}).

\bibitem[{\citenamefont{Moll and Rosencrans}(1990)}]{Moll-Rosencrans-1990}
\bibinfo{author}{\bibfnamefont{V.}~\bibnamefont{Moll}} \bibnamefont{and}
  \bibinfo{author}{\bibfnamefont{S.~I.} \bibnamefont{Rosencrans}},
  \bibinfo{journal}{SIAM J. Appl. Math.} \textbf{\bibinfo{volume}{50}},
  \bibinfo{pages}{1419} (\bibinfo{year}{1990}).

\bibitem[{\citenamefont{Idris and Biktashev}(2007)}]{Idris-Biktashev-2007}
\bibinfo{author}{\bibfnamefont{I.}~\bibnamefont{Idris}} \bibnamefont{and}
  \bibinfo{author}{\bibfnamefont{V.~N.} \bibnamefont{Biktashev}},
  \bibinfo{journal}{Phys. Rev. E} \textbf{\bibinfo{volume}{76}},
  \bibinfo{pages}{021906} (\bibinfo{year}{2007}).

\bibitem[{\citenamefont{Zel'dovich and Frank-Kamenetsky}(1938)}]{ZFK-1938}
\bibinfo{author}{\bibfnamefont{Y.~B.} \bibnamefont{Zel'dovich}}
  \bibnamefont{and} \bibinfo{author}{\bibfnamefont{D.~A.}
  \bibnamefont{Frank-Kamenetsky}}, \bibinfo{journal}{Doklady AN SSSR}
  \textbf{\bibinfo{volume}{19}}, \bibinfo{pages}{693} (\bibinfo{year}{1938}).

\bibitem[{\citenamefont{FitzHugh}(1961)}]{FitzHugh-1961}
\bibinfo{author}{\bibfnamefont{R.}~\bibnamefont{FitzHugh}},
  \bibinfo{journal}{Biophys. J.} \textbf{\bibinfo{volume}{1}},
  \bibinfo{pages}{445} (\bibinfo{year}{1961}).

\bibitem[{\citenamefont{Nagumo et~al.}(1962)\citenamefont{Nagumo, Arimoto, and
  Yoshizawa}}]{Nagumo-etal-1962}
\bibinfo{author}{\bibfnamefont{J.}~\bibnamefont{Nagumo}},
  \bibinfo{author}{\bibfnamefont{S.}~\bibnamefont{Arimoto}}, \bibnamefont{and}
  \bibinfo{author}{\bibfnamefont{S.}~\bibnamefont{Yoshizawa}},
  \bibinfo{journal}{Proc. IRE} \textbf{\bibinfo{volume}{50}},
  \bibinfo{pages}{2061} (\bibinfo{year}{1962}).

\bibitem[{\citenamefont{Gradshteyn and Rhyzhik}(2000)}]{Gradshteyn-Ryzhik2000}
\bibinfo{author}{\bibfnamefont{I.~S.} \bibnamefont{Gradshteyn}}
  \bibnamefont{and} \bibinfo{author}{\bibfnamefont{I.~M.}
  \bibnamefont{Rhyzhik}}, \emph{\bibinfo{title}{Tables of integrals, series,
  and products}} (\bibinfo{publisher}{Academic Press}, \bibinfo{year}{2000}),
  \bibinfo{edition}{sixth} ed.

\end{thebibliography}

\end{document}